\documentclass[runningheads]{llncs}
\usepackage{graphicx}
\usepackage{array,multirow}
\usepackage{float}
\usepackage[section]{placeins}

\begin{document}
\title{Studying the Effect of Digital Stain Separation of Histopathology Images on Image Search Performance \thanks{This research has been supported in part by the Natural Sciences and Engineering Research of Canada (NSERC) in the form of a Doctoral Scholarship (AKC) and a Discovery Grant (ERV).}}
\titlerunning{Digital Stain Separation of Histopathology Images for Image Search}
% If the paper title is too long for the running head, you can set
% an abbreviated paper title here
%
\author{Alison K. Cheeseman\inst{1} \and
Hamid R. Tizhoosh\inst{2}\and
Edward R. Vrscay\inst{1}}
\authorrunning{A.K. Cheeseman et al.}
% First names are abbreviated in the running head.
% If there are more than two authors, 'et al.' is used.
%
\institute{
Department of Applied Mathematics, Faculty of Mathematics,
University of Waterloo, Waterloo, Ontario, Canada N2L 3G1\\
\email{alison.cheeseman@uwaterloo.ca, ervrscay@uwaterloo.ca}
\and
Kimia Lab, University of Waterloo, Waterloo, Ontario, Canada N2L 3G1\\
\email{hamid.tizhoosh@uwaterloo.ca}
}
\maketitle              % typeset the header of the contribution
\begin{abstract}
Due to recent advances in technology, digitized histopathology images are now widely available for both clinical and research purposes. Accordingly, research into computerized image analysis algorithms for digital histopathology images has been progressing rapidly. In this work, we focus on image retrieval for digital histopathology images. Image retrieval algorithms can be used to find similar images and can assist pathologists in making quick and accurate diagnoses. Histopathology images are typically stained with dyes to highlight features of the tissue, and as such, an image analysis algorithm for histopathology should be able to process colour images and determine relevant information from the stain colours present. In this study, we are interested in the effect that stain separation into their individual stain components has on image search performance. To this end, we implement a basic k-nearest neighbours (kNN) search algorithm on histopathology images from two publicly available data sets (IDC and BreakHis) which are: a) converted to greyscale, b) digitally stain-separated and c) the original RGB colour images. The results of this study show that using H\&E separated images yields search accuracies within one or two percent of those obtained with original RGB images, and that superior performance is observed using the H\&E images in most scenarios we tested.

\keywords{Digital histopathology  \and Encoded Local Projections (ELP) \and Digital stain separation \and Digital image retrieval and classification.}
\end{abstract}
\section{Introduction}
Histopathology, the examination of tissue under a microscope to study biological structures related to disease manifestation, has traditionally been carried out manually by pathologists working in a lab. It is only in recent years that the technology has advanced to a point which allows for the rapid digitization and storage of whole slide images (WSIs). Consequently, digitized histopathology images are now widely available for both clinical and research purposes, and computerized image analysis for digital histopathology has quickly become an active area of research~\cite{ref_gurcan09,ref_madabhushi09}. In this paper, we focus specifically on content-based image retrieval (CBIR) for histopathology images, which involves finding images which share the same visual characteristics as the query image. The identification and analysis of similar images can assist pathologists in quickly and accurately obtaining a diagnosis by providing a baseline for comparison.

While most radiology images (X-ray, CT, etc.) are greyscale, histopathology images are typically stained with dyes to highlight certain features of the tissue. In order to properly use the relevant colour information, a WSI analysis system should be able to process colour images and determine relevant biological information from the presence of different stain colours. In a previous work~\cite{ref_cheeseman19}, we introduced a new frequency-based ELP (F-ELP) image descriptor for histopathology image search, which captures local frequency information and implemented this new descriptor on images which were separated into two colour channels based on chemical stain components using a digital stain separation algorithm. In~\cite{ref_cheeseman19}, we found that both the ELP and F-ELP descriptors saw improved search accuracy when applied to the stain-separated images, as opposed to single-channel greyscale images. In this paper, we focus on studying the effectiveness of digital stain separation of histopathology images for image retrieval applications using a number of common handcrafted image descriptors. We compare the results of image retrieval for the stain-separated images to the results of the same experiment conducted on both greyscale and colour (three-channel RGB) images. Experiments are conducted using two publicly available breast cancer histopathology data sets, IDC and BreakHis.

\section{Digital Stain Separation}
The most common staining protocol for histology slides involves two dyes, namely hematoxylin and eosin (H\&E). The hematoxylin stains cell nuclei blue, and eosin stains other structures varying shades of red and pink~\cite{ref_ruifrok01}. The colours which appear in a slide, and the size, shape and frequency at which they appear are all relevant factors a pathologist may assess when making a diagnosis. For this reason, we consider separating the input images into two stain components prior to the computation of an image descriptor.

In this paper, as in~\cite{ref_cheeseman19}, we adopt the stain separation method proposed in~\cite{ref_mccann14}, an extension of the wedge finding method from~\cite{ref_macenko09}. Unlike some previous methods for stain separation~\cite{ref_ruifrok01}, this method does not require any calibration or knowledge of the exact stain colours, instead it works by using the available image data to estimate an H\&E basis. Given that an image search algorithm should ultimately be applied to data from multiple sources, this is a desirable feature for the stain separation algorithm.

\begin{figure}[ht]
\centering
\includegraphics[width=0.95\textwidth]{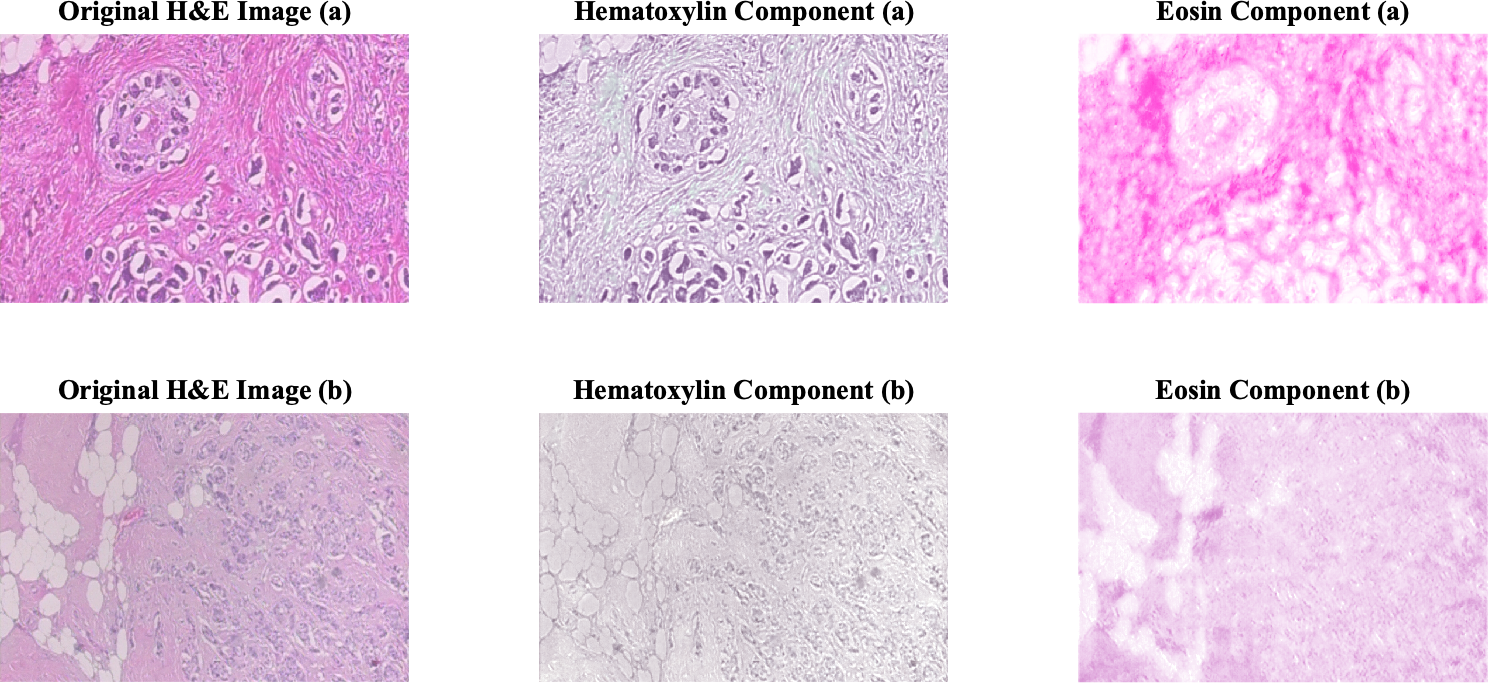}
\caption{Two sample images from the BreakHis dataset showing the resulting hematoxylin and eosin  components after applying the stain separation algorithm from~\cite{ref_mccann14}.} \label{fig_BH_stainseparated}
\end{figure}

Figure~\ref{fig_BH_stainseparated} shows two sample images from the BreakHis data set which have been separated in their hematoxylin and eosin components. We can see that the algorithm is able to effectively separate the two components of both images, even though the stains appear as noticeably different colours in each image.
\section{Proposed Study}
The main purpose of this study is to investigate the effectiveness of digital stain separation in isolation from the other parameters of the image search process. To this end, we implement a basic image search algorithm, k-Nearest Neighbours (kNN), using input images that are: a) converted to greyscale (one colour channel), b) separated into their H\&E components (two colour channels) by the method described above, and c) the original RGB colour images (three colour channels). For each colour channel of an input image having $N$ total channels, an image descriptor, $\mathbf{h}_i$ is computed, and concatenated to form the final descriptor $\mathbf{h} = [\mathbf{h}_1\ ...\ \mathbf{h}_N]$, which represents the entire image. We then implement the kNN algorithm for image search using a number of well-known distance functions to determine ``nearest" neighbours and four different image descriptors of varying lengths and properties as inputs to the algorithm. The image descriptors and distance functions used are described in the following sections.

\subsection{Image Descriptors}
Feature extraction, or the computation of compact image descriptors, is an important part of many image analysis tasks, including image retrieval. As such, there has been plenty of research over the years into the design of image descriptors for various imaging applications. Some of the most well known image descriptors include: local binary patterns (LBP)~\cite{ref_ahonen06}, the scale-invariant feature transform (SIFT)~\cite{ref_lowe04}, speeded-up robust features (SURF)~\cite{ref_bay08}, and histograms of oriented gradients (HOG)~\cite{ref_dalal05}. Most of these methods, including SIFT, SURF, and HOG, perform well in more traditional applications such as object detection or tracking and face recognition, but perform poorly compared to LBP for the retrieval and classification of histopathology images~\cite{ref_alhindi18,ref_tizhoosh18}. LBP, which is based on computing binary patterns in local regions of the image, is generally thought to be a better image descriptor for textures, which may explain its superior performance on high resolution histopathology images, which resemble textures more than natural images.

In our study, we implement four image descriptors, including the LBP descriptor, along with the Gabor filter-based GIST descriptor~\cite{ref_oliva01} which computes the spatial envelope of a scene, the ELP descriptor (encoded local projections) from~\cite{ref_tizhoosh18} which was designed with medical images in mind, and our proposed F-ELP descriptor from~\cite{ref_cheeseman19}, designed specifically to be a compact descriptor for histopathology images.

\subsection{Distance Functions for Image Search}
In any image search algorithm, it is important to properly define what makes two images similar. Typically, that means one must choose an appropriate distance function between image descriptors. Six different distance functions  were used in this study to determine the nearest neighbours for the kNN search, including the well-known $L_1$, $L_2$, cosine, correlation, and chi-squared metrics. We also consider the Hutchinson (also known as Monge-Kantorovich) distance~\cite{ref_mendivil17}, as it is thought to be a good measure of distance between histograms. In the finite one-dimensional case, the Hutchinson distance can be computed in linear-time using the method from~\cite{ref_molter91}.

\section{Data Sets \& Image Preprocessing}
In this study, we used two publicly available data sets containing breast cancer histopathology images: IDC and BreakHis.

\subsubsection{Invasive Ductal Carcinoma (IDC) Kaggle Data:} The IDC dataset consists of digitized breast cancer slides from 162 patients diagnosed with IDC at the University of Pennsylvania Hospital and the Cancer Institute of New Jersey~\cite{ref_cruz-roa14}. Each slide was digitized at 40x magnification and downsampled to a resolution of 4 $\mu$m/pixel.  The dataset provides each WSI split into patches of size 50px $\times$ 50px in RGB colour space. The supplied data was randomly split into three different subsets of 84 patients for training, 29 for validation and 49 test cases for final evaluation. Ground truth annotation regarding the presence of IDC in each patch was obtained by manual delineation of cancer regions performed by expert pathologists.
\begin{figure}[hb]
\centering
\includegraphics[width=0.99\textwidth]{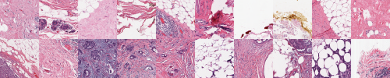}
\caption{Sample patches from the IDC data set. The top row shows negative examples (healthy tissue or non-invasive tumour tissue) and the bottom row shows positive examples (IDC tissue).} \label{fig_IDC_ims}
\end{figure}

Due to their small size, each individual image patch in the IDC data set may not contain both hematoxylin and eosin stains. Since the stain separation algorithm learns the stain colours from the data, both stains must be present in the image for accurate results. To ensure good performance on all image patches, we use the entire WSI to perform stain separation and then split the image back into the original patches to compute image descriptors. One further issue is that the stain separation algorithm used assumes that two (and only two) stain components (H\&E in our case) exist in the image. However, some images are observed to have significant discolouration, such as large dark patches, and the introduction of other colours not caused by H\&E staining. The prevalence of such artefacts negatively impacts the ability of the stain separation algorithm to provide good results for some patients, so we remove them by searching for images which have minimal variation in the RGB channels across the entire image. A total of 686 patches were flagged and removed from the total data set, all of which contain significant artefacts or discoloration.

\subsubsection{Breast Cancer Histopathology Database (Breakhis):} The Breast Cancer Histopathology Database (BreakHis)~\cite{ref_spanhol16} was built as a collaboration between researchers at the Federal University of Parana (UFPR) and the P\&D Laboratory - Pathological Anatomy and Cytopathology, in Parana, Brazil. To date, it contains 9,109 images of breast tumour tissue from 82 patients using four different magnification factors: 40$\times$, 100$\times$, 200$\times$, and 400$\times$. The images are provided in PNG format (3-channel RGB, 8-bit depth/channel) and are 700$\times$460 pixels. The data is divided into two classes, benign tumours and malignant tumours, with class labels provided by pathologists from the P\&D Laboratory. Within each class, further labelling is provided to indicate tumour types. The data set consists of four histologically distinct benign tumours and four malignant tumour types. These additional intra-class labels are not used in the current study.
\begin{figure}[ht]
\centering
\includegraphics[width=0.8\textwidth]{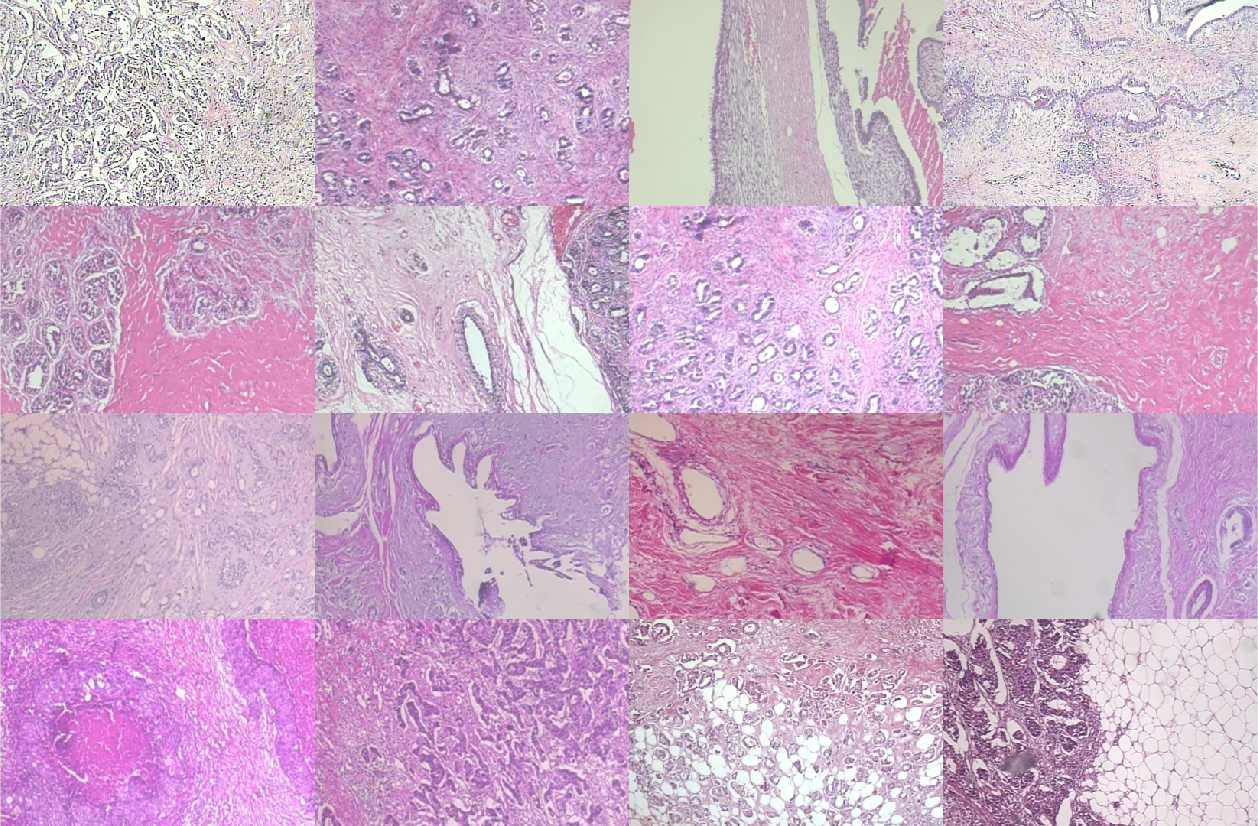}
\caption{Example of image patches from the BreakHis data set. The top two rows show examples of benign tumours and the bottom two rows show malignant tumours.} \label{fig_BH_ims}
\end{figure}

For this particular study, we use only the images taken at 40$\times$ magnification. This subset of the data contains 1,995 images, of which 652 are benign and 1,370 are malignant. Using code provided by the authors of~\cite{ref_spanhol16}, the data is split into a training (70\%) and testing (30\%) set with the condition that patients in the training set are not used for the testing set. The results presented in this paper are the average of five trials, using the five data folds from~\cite{ref_spanhol16}.

\subsection{Accuracy Calculations}
\subsubsection{IDC:}For consistency with previous works involving the IDC data, we use both the balanced accuracy (BAC) and F-measure (F1), which are defined as follows~\cite{ref_cruz-roa14}:
\begin{equation}
    \mbox{BAC} = \frac{\mbox{Sen}+\mbox{Spc}}{2}\mbox{, }F1 = \frac{2 \cdot \mbox{Pr} \cdot \mbox{Rc}}{\mbox{Pr}+\mbox{Rc}},
    \label{eq:BACF1}
\end{equation}
where Sen is sensitivity, Spc is specificity, Pr is precision and Rc, recall.

\subsubsection{BreakHis:}For the BreakHis data we compute patient scores and the global recognition rate, which were introduced in ~\cite{ref_spanhol16}. If we let $N_P$ be the number of images of patient $P$ and $N_{\mbox{rec}}$ be the number of images of patient $P$ that are correctly classified, then the patient score for patient $P$ is defined as
\begin{equation}
    \mbox{Patient Score} = \frac{N_{\mbox{rec}}}{N_P}
\end{equation}
and the global recognition rate (GRR) as
\begin{equation}
    \mbox{Global Recognition Rate} = \frac{\sum \mbox{Patient scores}}{\mbox{Total number of patients}}.
\end{equation}
In addition to the global recognition rate we also compute the balanced accuracy as defined above in~(\ref{eq:BACF1}).

\section{Experiment}
In order to evaluate image retrieval performance we implement the kNN algorithm in MATLAB with each set of image descriptors as inputs. The kNN algorithm searches through the training data partition and classifies each image in the test set based on the class of its $k$ nearest neighbours. Since there is no exact metric to quantitatively evaluate image retrieval performance, we measure the accuracy of classification using kNN. In this work, we test the kNN algorithm using three different values for $k$ ($k=1,5$ and $15$).
\begin{table}[ht]
\centering
\caption{A list of image descriptors used in this study and the corresponding number of features computed (i.e. the length of the feature vector).}\label{tab_descriptors}
\begin{tabular}{|c|c|c|c|}
\hline
\multirow{2}{*}{Descriptor} & \multicolumn{3}{c|}{Number of Features}\\\cline{2-4}
& Greyscale & H\&E Stains & RGB Image \\\hline
ELP	    	    & 1024 & 2048 & 3072\\
GIST 			& 512 & 1024 & 1536 \\
F-ELP       	& 32 & 64 & 96 \\
LBP	    		& 18 & 36 & 54\\
\hline
\end{tabular}
\end{table} 

Table~\ref{tab_descriptors} lists the image descriptors used and their respective lengths on each set of input colour channels. We can see that as we increase the number of input colour channels from one to three, the length of the feature vectors increases. Given that the computation time for the kNN search algorithm has linear dependency on feature vector length~\cite{ref_matlabknn}, it is clear that for a fixed image descriptor, it is desirable to use fewer colour channels, so long as the overall search performance does not suffer significantly.

The length of each image descriptor is dependent on certain parameters of the algorithm. In this work the following parameters are used: the ELP and F-ELP descriptors are implemented with a window size of $w=9$, the GIST descriptor, by default, divides the image into a $4 \times 4$ grid and uses a filter bank of $32$ Gabor filters, and the LBP descriptor is computed with a radius of $R=2$ and $P=16$ neighbouring pixels. As a result, we have a wide variety of descriptor lengths, from the very short LBP descriptor to the long ELP histogram.

\subsection{Comparing Input Image Colour Channels}
In this section, we present results which compare the image search performance using greyscale, H\&E stain-separated, and RGB images as inputs. For each descriptor, and each set of input colour channels, the best accuracy, taken over all distance functions, is presented. It should be noted here that there are some slight discrepancies between the results presented here for the IDC data set and those in our previous work~\cite{ref_cheeseman19}. This is due to a small error which was found in the code which slightly changes the numerical results, but does not change the overall conclusions of the previous study.

\subsubsection{IDC:} 
\begin{table}[ht]
  \centering
  \caption{The best KNN search ($k=1$) accuracy for the IDC dataset taken over all distance functions. The top result in each column is highlighted in bold.}
    \begin{tabular}{|c|c|c|c|c|c|c|c|c|}
    \hline
    \multirow{2}{*}{Colour Channels} & \multicolumn{2}{c|}{ELP} & \multicolumn{2}{c|}{GIST} & \multicolumn{2}{c|}{F-ELP} & \multicolumn{2}{c|}{LBP} \\\cline{2-9}
    & F1   & BAC   & F1   & BAC   & F1   & BAC   & F1   & BAC \\\hline
    Greyscale & 0.3987 & 0.5905 & 0.5086 & 0.6500 & 0.4183 & 0.6023 & 0.4625 & 0.6280\\
    H\&E Stains & \textbf{0.4528} & \textbf{0.6235} & \textbf{0.5549} & \textbf{0.6923} & \textbf{0.5565} & \textbf{0.6932} & 0.5860 & 0.7130\\
    RGB Image & 0.4504 & 0.6219 & 0.5513 & 0.6890 & 0.5419 & 0.6836 & \textbf{0.5926} & \textbf{0.7187}\\
    \hline
    \end{tabular}
  \label{tab:IDCSUMK1}
\end{table}
\begin{table}[ht]
  \centering
  \caption{The best KNN search ($k=5$) accuracy for the IDC dataset taken over all distance functions. The top result in each column is highlighted in bold.}
    \begin{tabular}{|c|c|c|c|c|c|c|c|c|}
    \hline
    \multirow{2}{*}{Colour Channels} & \multicolumn{2}{c|}{ELP} & \multicolumn{2}{c|}{GIST} & \multicolumn{2}{c|}{F-ELP} & \multicolumn{2}{c|}{LBP} \\\cline{2-9}
    & F1   & BAC   & F1   & BAC   & F1   & BAC   & F1   & BAC \\\hline
    Greyscale & 0.4001 & 0.6080 & 0.5598 & 0.6968 & 0.4338 & 0.6208 & 0.5124 & 0.6645\\
    H\&E Stains & \textbf{0.4880} & \textbf{0.6531} & \textbf{0.6052} & \textbf{0.7356} & \textbf{0.6303} & \textbf{0.7406} & 0.6618 & 0.7614\\
    RGB Image & 0.4832 & 0.6504 & 0.5918 & 0.7270 & 0.6028 & 0.7222 & \textbf{0.6785} & \textbf{0.7750}\\
    \hline
    \end{tabular}
  \label{tab:IDCSUMK5}
\end{table}
\begin{table}[ht]
  \centering
  \caption{The best KNN search ($k=15$) accuracy for the IDC dataset taken over all distance functions. The top result in each column is highlighted in bold.}
    \begin{tabular}{|c|c|c|c|c|c|c|c|c|}
    \hline
    \multirow{2}{*}{Colour Channels} & \multicolumn{2}{c|}{ELP} & \multicolumn{2}{c|}{GIST} & \multicolumn{2}{c|}{F-ELP} & \multicolumn{2}{c|}{LBP} \\\cline{2-9}
    & F1   & BAC   & F1   & BAC   & F1   & BAC   & F1   & BAC \\\hline
    Greyscale & 0.3839 & 0.6069 & 0.5910 & 0.7207 & 0.4218 & 0.6199 & 0.5396 & 0.6825 \\
    H\&E Stains & \textbf{0.4943} & \textbf{0.6589} & \textbf{0.6283} & \textbf{0.7570} & \textbf{0.6697} & \textbf{0.7665} & 0.6887 & 0.7780 \\
    RGB Image & 0.4912 & 0.6569 & 0.6147 & 0.7448 & 0.6404 & 0.7462 & \textbf{0.7125} & \textbf{0.7972} \\
    \hline
    \end{tabular}
  \label{tab:IDCSUMK15}
\end{table}
Tables~\ref{tab:IDCSUMK1},~\ref{tab:IDCSUMK5} and~\ref{tab:IDCSUMK15} show the results for the IDC data set for kNN search with $k=1,5$ and $15$, respectively. As expected, since coloured images contain relevant information which may be lost when converted to greyscale, we observe that using either the H\&E stain separated image or the total RGB image is always an improvement over using the greyscale image. A more interesting comparison comes from looking at the bottom two rows of the tables, comparing the H\&E images to the RGB images. We see that generally the F1 scores and balanced accuracies are similar (within one or two percent) for both H\&E and RGB images. For all descriptors besides LBP, we actually observe an improved performance using the H\&E image over RGB, despite the fact that the input image has less colour channels, and thus the feature vector is shorter.

We also, not surprisingly, observe that as $k$ is increased, the search performance tends to improve, although the jump from $k=1$ to $5$ is quite a bit larger than the jump from $5$ to $15$. This may indicate that $k=15$ is near an optimal value for $k$. We also see that for this particular data set, the LBP descriptor gives the highest accuracy, and the ELP descriptor performs the worst.

\subsubsection{BreakHis:} Similarly, for the BreakHis data set, Tables~\ref{tab:BHSUMK1},~\ref{tab:BHSUMK5} and~\ref{tab:BHSUMK15} show the best global recognition rates and balanced accuracies for each image descriptor and set of input colour channels. Once again, we observe a general increase in search accuracy when using more than one input colour channel (H\&E or RGB) as compared to the greyscale images. In many cases, in addition to the decreased computational cost of using fewer colour channels, for the BreakHis data set we see that there is another benefit to using stain separated images, which is a significant improvement in search performance. 
\begin{table}[ht]
  \centering
  \caption{The best KNN search ($k=1$) accuracy for the BreakHis dataset taken over all distance functions. The top result in each column is highlighted in bold.}
    \begin{tabular}{|c|c|c|c|c|c|c|c|c|}
    \hline
    \multirow{2}{*}{Colour Channels} & \multicolumn{2}{c|}{ELP} & \multicolumn{2}{c|}{GIST} & \multicolumn{2}{c|}{F-ELP} & \multicolumn{2}{c|}{LBP} \\\cline{2-9}
    & GRR   & BAC   & GRR   & BAC   & GRR   & BAC   & GRR   & BAC \\\hline
    Greyscale & 0.6584 & 0.5812 & 0.6589 & 0.5602 & 0.6534 & 0.5577 & 0.6874 & 0.6212 \\
    H\&E Stains & \textbf{0.7532} & \textbf{0.7047} & \textbf{0.7083} & \textbf{0.6456} & \textbf{0.7433} & \textbf{0.6823} & \textbf{0.7397} & \textbf{0.6903}\\
    RGB Image & 0.6604 & 0.5971 & 0.6578 & 0.5787 & 0.7358 & 0.6812 & 0.6689 & 0.6115\\
    \hline
    \end{tabular}
  \label{tab:BHSUMK1}
\end{table}

As before, we see that search performance tends to increase with increasing $k$, but does seem to level off around $k=15$. Unlike our previous results on the IDC data set, we see here that the best search performance occurs using the ELP descriptor and F-ELP descriptors, while the GIST descriptor fares the worst on the BreakHis data. Due to the higher intra-class variation of the BreakHis data (multiple tumour types for benign and malignant classes), it makes sense that the balanced accuracies are generally lower on this data set.
\begin{table}[ht]
  \centering
  \caption{The best KNN search ($k=5$) accuracy for the BreakHis dataset taken over all distance functions. The top result in each column is highlighted in bold.}
    \begin{tabular}{|c|c|c|c|c|c|c|c|c|}
    \hline
    \multirow{2}{*}{Colour Channels} & \multicolumn{2}{c|}{ELP} & \multicolumn{2}{c|}{GIST} & \multicolumn{2}{c|}{F-ELP} & \multicolumn{2}{c|}{LBP} \\\cline{2-9}
    & GRR   & BAC   & GRR   & BAC   & GRR   & BAC   & GRR   & BAC \\\hline
    Greyscale & 0.6838 & 0.5844 & 0.6940 & 0.5808 & 0.6802 & 0.5637 & 0.6952 & 0.6049 \\
    H\&E Stains & \textbf{0.7602} & \textbf{0.7078} & \textbf{0.7340} & \textbf{0.6324} & \textbf{0.7557} & \textbf{0.6915} & \textbf{0.7323} & \textbf{0.6662} \\
    RGB Image & 0.6744 & 0.5931 & 0.6884 & 0.6113 & 0.7480 & 0.6865 & 0.6977 & 0.6335 \\
    \hline
    \end{tabular}
  \label{tab:BHSUMK5}
\end{table}
\begin{table}[ht]
  \centering
  \caption{The best KNN search ($k=15$) accuracy for the BreakHis dataset taken over all distance functions. The top result in each column is highlighted in bold.}
    \begin{tabular}{|c|c|c|c|c|c|c|c|c|}
    \hline
    \multirow{2}{*}{Colour Channels} & \multicolumn{2}{c|}{ELP} & \multicolumn{2}{c|}{GIST} & \multicolumn{2}{c|}{F-ELP} & \multicolumn{2}{c|}{LBP} \\\cline{2-9}
    & GRR   & BAC   & GRR   & BAC   & GRR   & BAC   & GRR   & BAC \\\hline
    Greyscale & 0.7085 & 0.5898 & 0.7128 & 0.5744 & 0.6957 & 0.5616 & 0.7051 & 0.5979 \\
    H\&E Stains & \textbf{0.7660} & \textbf{0.7060} & \textbf{0.7406} & \textbf{0.6286} & 0.7737 & \textbf{0.7033} & \textbf{0.7564} & \textbf{0.6885} \\
    RGB Image & 0.6892 & 0.6008 & 0.7068 & 0.6090 & \textbf{0.7749} & \textbf{0.7033} & 0.7023 & 0.6219 \\
    \hline
    \end{tabular}
  \label{tab:BHSUMK15}
\end{table}
\FloatBarrier

\subsection{Comparing Distance Functions} 
In this section, we consider the effect that the choice of distance function has on image search performance for each image descriptor. To do so, we introduce a ranking of each distance function, based on the balanced accuracy result. For a given image descriptor and choice of input colour channels (greyscale, H\&E, or RGB) we rank each distance function based on the resulting balanced accuracy as a percentage of the maximum balanced accuracy for that particular search trial. Results presented here show the averaged distance ranking for each distance and each descriptor, averaged over the choice of input colour channels, and the choice of $k$ for the kNN search algorithm. We present results only for the balanced accuracy, as the results for the F1 measure (IDC data) and global recognition rate (BreakHis data) follow similar trends.

\subsubsection{IDC:} 
\begin{figure}[ht]
\centering
\includegraphics[width=0.62\textwidth]{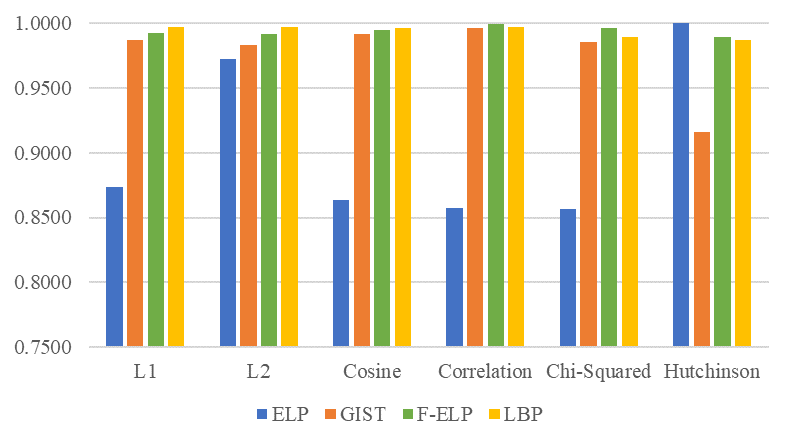}
\caption{A comparison of the average ranking of distance functions for each image descriptor on the IDC data set.} \label{fig_IDC_distchart}
\end{figure}

Figure~\ref{fig_IDC_distchart} shows the average ranking of all six distance functions tested for each image descriptor on the IDC data. We observe that, in general, the variation in search accuracy caused by the choice of distance function is relatively low, with the exception of the ELP descriptor, where a noticeable variation can be seen. For all but the ELP descriptor, it would be difficult to pinpoint which distance function is the ``best" choice. In particular, for the F-ELP and LBP descriptors, the variation in performance across distance functions is almost non-existent. 

\subsubsection{BreakHis:} 
\begin{figure}[ht]
\centering
\includegraphics[width=0.62\textwidth]{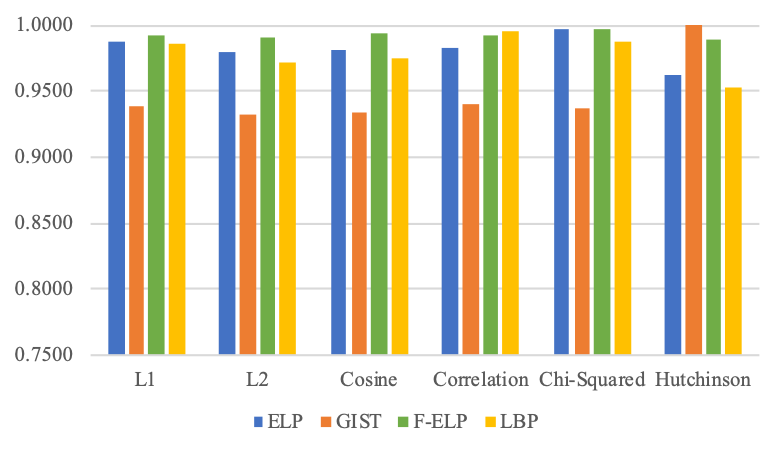}
\caption{A comparison of the average ranking of distance functions for each image descriptor on the BreakHis data set.} \label{fig_BH_distchart}
\end{figure}

In Figure~\ref{fig_BH_distchart} we show the average rankings of each distance function on the BreakHis data. Once again, the overall change in the accuracy as a result of the choice of distance function is surprisingly low for all descriptors. As with the IDC data, we see the most noticeable variation in performance with the ELP and GIST descriptors.
The results for both the IDC and BreakHis data sets do not give any indication that one distance function is necessarily superior for image search even for a fixed image descriptor. Over many tests, we see only one scenario (the ELP descriptor applied to the IDC data) where the choice of distance function significantly impacts the results. Further testing on more data is of course possible, however for good performance and generalisation to unknown data, it would seem that the best choice is simply to use the distance function which can be computed most efficiently.

\section{Conclusion}

In this paper, we have investigated the effect of using digitally stain separated images, as compared to greyscale and RGB, for image retrieval applications. Our results are obtained through testing on two data sets containing breast cancer histopathology images. We find that separating images into their H\&E stain components leads to a significant increase in search performance over simply using the greyscale images, as expected. More interestingly, we find that using H\&E separated images yields search accuracies within one or two percent of those obtained with the original RGB images, despite the fact that the H\&E images have only two colour channels. In fact, superior performance is observed using the H\&E images in most tested scenarios. Given the improved computation speed afforded by using fewer image channels, it is reasonable to conclude that using H\&E stain separated images is preferable to using the overall RGB images for image search. As well, ELP appears to benefit from investigations on choosing the distance metric, a factor that should be considered when using this descriptor.
%
% ---- Bibliography ----
%
\bibliographystyle{splncs04}
\bibliography{references}
\end{document}